\documentclass[english,aps,prper,reprint,showpacs,titlepage,longbibliography,nofootinbib]{revtex4-2}   

\usepackage[T1]{fontenc}	
\usepackage[latin9]{inputenc}	
\usepackage{geometry}		
\usepackage{graphicx}
\usepackage[above,below]{placeins}	
\usepackage{times}

\usepackage{hyperref}  
\hypersetup{colorlinks=true,urlcolor=blue,citecolor=blue,linkcolor=blue}   
\usepackage{enumitem}          
\setlist{nosep}                 

\usepackage{csquotes} 

\usepackage{xcolor}

\begin{document}

\newcommand{\mysection}[1]{
\vspace{-1 em}
\section{#1}
\vspace{-1.2 em}
}

\newcommand{\mysubsection}[1]{
\vspace{-1 em}
\subsection{#1}
\vspace{-1.2 em}
}

\newcommand{\mysectionNoNumber}[1]{
\vspace{-1 em}
\section*{#1}
\vspace{-1.2 em}
}

\newcommand{\checkThis}[1]{\textcolor{red}{#1}}

\begin{titlepage}

  \title{What is interdisciplinarity? Views from students and professors in a non-major intro physics course}

  \author{Ian Descamps} 
  \email[Please address correspondence to ]{ian.descamps@pomona.edu}
  \affiliation{Department of Physics, Pomona College, Claremont, CA 91711}
  \author{Thomas Moore}
  \affiliation{Department of Physics, Pomona College, Claremont, CA 91711 }
  \author{Benjamin Pollard}
  \affiliation{Department of Physics, University of Colorado Boulder, Boulder, CO 80309, USA}
  \affiliation{JILA, National Institute of Standards and Technology, Boulder, CO 80309, USA }


  \begin{abstract}
We present an investigation into the interdisciplinary role of physics in a physics-for-non-physicists course at Pomona College.
This work is guided by prior research into introductory physics for life-science (IPLS) courses, but attends to significant differences in the scope and context of this course.
We interviewed enrolled students, physics professors, and professors from non-physics disciplines to explore the function of this course and the role of physics in the education of non-physics-science students.
Interviews were audio recorded and transcribed, then analyzed to identify emergent themes.
These themes outline the authentic physics, including content knowledge and other, broader learning objectives, that play an important and distinct role in the science education of enrolled students.
Stakeholders generally align in their emphasis of interdisciplinary relevance with some divergence in the specific articulation of that idea.
The differences can be understood through the stakeholders' distinct areas of expertise, with non-physics professors expressing value through relevance to their discipline and physics professors focusing on essential aspects of physics.

\clearpage
  \end{abstract}

  \maketitle
  
\end{titlepage}

\mysection{Introduction}
The introductory physics for life-science (IPLS) course is not a new phenomenon in physics departments.
Still, the last decade has seen significant, renewed research activity focused on such courses \cite{redish_reinventing_2009,crouch_physics_2010,meredith_reinventing_2013,redish_nexus/physics:_2014,crouch_introductory_2014,geller_sources_2018}, responding to national calls for more effective physics education for biologists and pre-medical students \cite{national_research_council_bio2010:_2003, noauthor_scientific_nodate, vision_and_change}.
The scientific skills, quantitative literacy, and interdisciplinary coherence that might ideally result from effective interdisciplinary physics education \cite{national_research_council_bio2010:_2003, noauthor_scientific_nodate, vision_and_change} are among the competencies that that define the role of physics on the Medical College Admissions Test (MCAT) \cite{hilborn_physics_2014} and are often listed as goals for IPLS courses \cite{redish_nexus/physics:_2014,crouch_introductory_2014}.

Specific interdisciplinary physics education, typified in Physics Education Research (PER) by IPLS courses, functions to meet the unique needs of non-physics students in physics courses.
Such students are likely to only take one or two physics courses, have less familiarity with math, and be preparing for expertise in scientific fields outside of physics \cite{meredith_reinventing_2013}.
These students have a different relationship to the physics discipline and, thus, to their physics courses relative to physics majors.

The course studied here, General Physics, is one example of such interdisciplinary physics education:  the students are predominantly non-physics science students and they most often take this physics course to support and advance their studies in a different discipline.
In this paper, we aim to answer the following research questions: How are the goals of the different stakeholders (students, physics faculty, non-physics faculty/departments) aligned? How are they distinct? 
Our work seeks to understand the role of this course, with particular attention paid to the interdisciplinarity at its foundation.
Prior IPLS research motivates our careful consideration of non-physics perspectives and stakeholders, but, as described below, our investigation differs in both structure and content from that work.

Recent IPLS work has focused on designing courses that attend to the specific context of their interdisciplinary population \cite{redish_nexus/physics:_2014, crouch_introductory_2014}.
The University of Maryland course prioritizes epistemology and interdisciplinary coherence \cite{redish_nexus/physics:_2014}. This focus stems from studies involving students, \cite{watkins_context_2013, gouvea_framework_2013}, discussions with biology faculty and biology education faculty, and prior research \cite{redish_reinventing_2009}.
At the University of Minnesota and Swarthmore College, surveys of and workshops with life-science faculty shaped their course goals and curricula \cite{crouch_introductory_2014}.
In these course transformations, input and feedback from non-physics stakeholders led to key interventions to better align the IPLS courses with the needs of its population; for example, all the previously cited courses restructured their discussions of energy so students would develop a cohesive view of this concept across several science disciplines \cite{crouch_introductory_2014, redish_nexus/physics:_2014}

Our work builds on these recent IPLS projects by drawing on other studies concerning the process of transforming courses more generally.
For example, in Chasteen et al. \cite{chasteen_thoughtful_nodate} researchers at the University of Colorado and the University of British Columbia Science Education Initiative (SEI) operationalized their approach to research-based course transformations. 
They detail seven key features that encompass the preparation, enactment, assessment, and sustainability of an effective course transformation. 
For preparation, they outline three features:  project scope, learning goals, and documenting student thinking.	

While Chasteen et al. provide suggestions for research to support their various preparation steps, Lewandowski et al. \cite{lewandowski_studying_2015} provide a concrete example that involves consulting different stakeholders to identify learning goals.  
For their upper-division electronics course, the authors interviewed both faculty and graduate student researchers in order to outline goals for their course. 
They note that their inclusion of graduate students in the interviews facilitated a better alignment of their research with a major function of their course -- to prepare students for a research career. 
This idea naturally extends to IPLS, where research has documented how the expertise of non-physics stakeholders generates crucial interventions \cite{redish_reinventing_2009, redish_nexus/physics:_2014, crouch_introductory_2014,watkins_disciplinary_2012}.

By drawing on these more general models of course transformation, we extend recent IPLS research in both structure and content.
Regarding content, here we study 
non-physics faculty and students across a broader range of disciplines, including geology, environmental analysis, and chemistry. 
Investigations of the views of such a broader range of stakeholders, in conjunction with the views of physics faculty, can yield a more complete understanding of the role of physics in the education of non-physics-science-students.
Regarding structure, we conduct formal interviews and subsequent analysis specifically focused on the role of interdisciplinarity across this range of stakeholders.
Similar questions have been addressed using qualitative methods in prior IPLS work, but none with the specific focus of comparing perspectives around interdisciplinarity between these three stakeholder groups.
With such an approach, we aim to uncover deeper insight into the perspectives and motivations of physics faculty and a broad range of non-physics faculty and students.



As stated above, we aim to answer the following research questions: How are the goals of the different stakeholders (students, physics faculty, non-physics faculty/departments) aligned? How are they distinct? 
In this paper, we highlight the importance of additional methods for incorporating and utilizing non-physics stakeholders in the context of course reform, illustrate key steps in an interdisciplinary course reformation process, and offer unique variations on the approaches to IPLS education described elsewhere. 

In Sections \ref{backgroundSection} and \ref{contextSection}, we elaborate further on the context of Pomona College and the physics course studied. 
Then, in Section \ref{methodsSection}, we outline the interview methodologies and thematic analysis techniques that we employ in this work.
Section \ref{resultsSection} presents the perspectives from different invested parties -- enrolled students, non-physics faculty members, and physics faculty -- in the form of quotes from our interviews. 
Finally, in Section \ref{discussionSection}, we discuss the implications of these perspectives and, in Section \ref{conclusionSection}, highlight future research possibilities and offer conclusions.

\section{Background}\label{backgroundSection}
Over a decade ago, the Pomona College Physics faculty recognized that the non-physics population of students in the ``one-size-fits-all'' introductory course offered at the time had a significantly different relationship to the physics material and to the discipline than the potential physics majors. 
Thus, they created General Physics to more effectively meet the needs of an interdisciplinary science population, similar to the IPLS courses cited above. 
While current faculty recall that the course was created without direct inspiration from the research referenced above, the department is currently interested in understanding the unique and innovative aspects that the course offers, and in studying the experiences of people involved with the course in order to improve it. 
Accordingly, authors ID and TM have started the first formal research project around General Physics at Pomona College, with the overall goal of providing specific guidance to the Pomona College Physics Department about this course.

General Physics serves an interdisciplinary population of students who come from a wide range of mostly science disciplines, not just the life sciences. 
Therefore, as IPLS has been defined in prior research \cite{redish_nexus/physics:_2014}, General Physics is \textbf{not} an IPLS course. 
Nonetheless, the research above provides the backdrop for our work here in that it models course reformation with interdisciplinary considerations. 
More so than the specific changes made for their IPLS courses, the process detailed in the citations above have guided this research. 
In fact, Crouch and Heller explicitly designed their paper to outline a reformation process that others can adapt to their own local context \cite{crouch_introductory_2014}. 
We do this by adapting the work of Lewandowki et al. \cite{lewandowski_studying_2015} to our IPLS-adjacent setting.
To the best of our knowledge, our approach is novel for IPLS research.
Furthermore, our work presents new content because our interview subjects include a broader interdisciplinary population, 
even as we compare our findings to prior IPLS research.

\mysection{Course context}\label{contextSection}
This work centers around an introductory calculus-based physics course at Pomona College. 
Pomona College is a private, selective, arts-and-science focused college in southern California with primarily full-time, four-year students \cite{CarnegieNote}. 
The course, General Physics, is the \textit{de facto} introductory physics course for life-science and pre-health students at Pomona College, although it also serves other science students as well. 
For example, General Physics is a pre-requisite for the upper division Physical Chemistry course. 
Additionally, students from the Geology and Environmental Analysis departments and those with relevant academic interests also take this course, though to a significantly smaller degree.
Regardless of major, most of the students in General Physics are expected to take two semesters of introductory physics, either as a formal requirement or via an informal recommendation for their academic path. 
	
General Physics consistently has a large enrollment, relative to other courses offered at Pomona College. 
In Fall 2018, when this study took place, 53 students took the course, which was split into two sections. 
Each section had a significantly larger enrollment than the average course at Pomona College, which is 15 students.
Furthermore, General Physics had a larger enrollment than the introductory physics course aimed at physics majors.

During this study, General Physics was taught in a traditional lecture format, with a laboratory component and weekly "mentor" sessions (optional sessions that are more focused on homework and run by undergraduate physics students).
Lectures were largely not interactive, although many students reported in interviews that they were engaging. 
The content of the course predominantly aligned with traditional introductory courses:  the first semester focused on forces, Newtonian mechanics, and fluids; the second semester largely explored electricity and magnetism, with brief units on waves and light, and thermodynamics and energy. 

While not the focus of this work, the course also includes a regular lab component taught by other physics professors in the department, which follows the content throughout the semester.
These labs mirror traditional introductory physics lab courses in most respects, with results often coming from a regression analysis and compared to known values, though some activities have a more open-ended nature and focus on modeling and interpreting data.
Students' lab work is evaluated through lab notebook checks and brief quizzes at the end of each session.

Self-reported demographic information about the students in General Physics was collected via an online survey in Fall 2018, and is shown in Table \ref{demographicTable}.
37 students answered questions about their year, major, race/ethnicity, and gender in open-response format.
Some students specified particular identities within the pan-ethnic Asian-Pacific Islander grouping, but we do not report the individual Asian identities here to protect the anonymity of the respondents.
Though this information is incomplete, General Physics has a more diverse population than the Physics Department at large \cite{APSStatsNote}.
\begin{table}
\caption{Self-reported year, Discipline,  race/ethnicity, and gender of students in General Physics in Fall 2018. The "Other" category for students' discipline includes, but is not limited to, psychology, geology, environmental analysis, and undecided. \label{demographicTable}}
\begin{center}
 \begin{tabular}{|l c|} 
 \hline
 \multicolumn{2}{|l|}{\textbf{Year}} \\
 \hline
 First \& Second Year  & 27  \\ 
 \hline
 Third Year & 7 \\ 
 \hline
 Fourth Year & 3 \\
 \hline\hline
 \multicolumn{2}{|l|}{\textbf{Discipline}} \\
 \hline
 Molecular Biology & 13 \\
 \hline
 Other & 11 \\
 \hline
 Chemistry & 8 \\
 \hline
 Biology & 5 \\
 \hline
 Neuroscience & 4 \\
  \hline\hline
  \multicolumn{2}{|l|}{\textbf{Race/Ethnicity}} \\
 \hline
 Asian-Pacific Islander/Asian-American & 16 \\
 \hline
 White & 13 \\
 \hline
 Mixed Race & 8 \\
 \hline
 Black/African/African-American & 5 \\
 \hline
 Latinx/Hispanic & 5 \\
  \hline\hline
  \multicolumn{2}{|l|}{\textbf{Gender}} \\
  \hline
 Female & 25 \\
 \hline
 Male & 12 \\
 \hline
\end{tabular}
\end{center}
\end{table}

 

\mysection{Methods}\label{methodsSection}
To investigate views around interdiciplinarity in General Physics, the first author (ID) conducted interviews with individuals from three populations connected to the course: students, course professors, and professors in departments other than Physics.
All interviews were semi-structured, drawing from an interview protocol created beforehand by the research team. 
The interview questions with faculty focused on the interdisciplinary role of physics and how General Physics might fit into that picture. 
The interview questions with students focused on their experience with this course. 

First, in the Fall of 2018, ID interviewed six tenured non-physics professors:  two biology professors, two chemistry professors, and two neuroscience professors.
Five professors are associated with the Molecular Biology program, and three are associated with the pre-health program at Pomona College.
Three were men and three were women. 
Interview participants were contacted by email based on the recommendations of department chairs, previous interview subjects, or author TM.
Department chairs were specifically asked who would be well suited to speak about interdisciplinary connections between physics and their department.
Professors associated with the Pomona College pre-health program were also sought for interviews.

After these interviews, ID interviewed 13 students in the course during Fall 2018. 
ID sent an email message and made an announcement during a lecture period explaining the project and asking for volunteers.
Their participation was compensated financially. 
This population consisted of 10 women and 3 men; 7 identified as Asian-Pacific Islander/Asian-American (including those who specified particular identities within Asia), 5 were underrepresented minority students \cite{URMNote},
3 identified as mixed race, and 3 students were white; 9 were second-years and 4 were third-year or fourth-years; 6 students specified their discipline as molecular biology, 2 as biology, 1 as neuroscience, 1 as chemistry, and 4 students specified disciplines categorized as other.

After the end of Fall 2018, ID interviewed two physics professors who had recently taught General Physics. 
The interviews with Physics faculty sought to more formally clarify instructors' views on the role of this course, the interdisciplinary role of physics, and how they approached teaching such a unique course.
These professors were aware of this work previously, having consulted with ID and having attended presentations by ID in the context of a related senior research capstone project.


All interviews were audio recorded and later transcribed for analysis.
Once an initial draft of this paper was completed, ID member checked the quotes and analysis with the participants; ID asked each participant if the quotes and corresponding analysis aligned with their overall sentiments.
Nearly all participants responded, and those who did affirmed the validity of the data presented here.

Our overall analysis approach aligns with the ``progressive refinement of hypotheses'' process described by Engle, Connant, and Greeno \cite{Engle2007}, and the phenomenographic methods used by Irving and Sayre \cite{Irving2015}. 
This work represents an initial phase of the ''progressive refinement of hypotheses'' process, wherein researchers collect relevant data around a general research question in order to inform more specific hypotheses to test and questions to answer.
In conducting this preliminary analysis, we also draw on ideas from ``constant comparison'' methodologies, as described and employed by Little et al. \cite{Little2019}, to refine the concepts and relationships in our analysis.
In this work, we focus on the analysis of an individual researcher, ID, who had a close relationship with the data, as the first step of the methodology presented in ref. \cite{Irving2015}. 

For analysis, an inductive thematic analysis method was used to organize and categorize quotations. 
Thematic categories were developed and defined on an iterative basis, and transcripts underwent several rounds of review to ensure completeness. 
The process focused on the content of an interviewee's responses and sought to make themes readily identifiable through categorization. 
Themes were identified and analyzed through examination of quotes associated with specific categories and/or with several categories.

After ID performed this thematic analysis, he selected eight quotes from the interviews, spanning each of the three themes presented below, which BP then categorized independently without knowledge of which categories ID had assigned to which quotes. 
Subsequent comparison of the categories selected by BP and ID for these quotes showed perfect agreement between the two researchers.
Moreover, subsequent discussion between those authors around these quotes further refined our understanding of the categories and the relationships between them.

\mysection{Results}\label{resultsSection}
In this section we examine responses that summarize and represent the perspectives of the three subject populations with regards to our investigation into learning goals for this course. 
In order to distill trends within and across the different stakeholders that present pertinent information about the interdisciplinary role and function of physics at Pomona College, we examine quotes associated with the following categories: 

\begin{displayquote}
\textbf{Interdisciplinary Relationship.} This category applies whenever the interviewee talks specifically about interdisciplinary aspects of General Physics, or about how physics (as a discipline) relates to other disciplines.

\textbf{Goals.} This category applies whenever the interviewee discusses potential learning goals or outcomes related to enrollment in General Physics and the study of physics. Examples may look like an explicit statement of what they hope to gain from this course but also like what they envision as success in the course or expressions of how students will use the physics learned in this course.

\textbf{Course Role/Function.} This category captures statements about the function of this course in the broader undergraduate experience. Examples may look like explanations for why an interviewee is taking the course or how the course fits into the curriculum of an academic major. 
\end{displayquote}

\mysubsection{Interdisciplinary Relationship}
Both physics and non-physics professors spoke clearly and specifically about the relationship(s) between physics and other disciplines.
Physics professors outlined what they viewed as the central aspects of physics as a scientific discipline, and linked those ideas to physics' interdisciplinary value.
\begin{displayquote}I think in the sort of hierarchy of the reduction-to-the-absurd approach to science, that physics lies really underneath all of everything. \textbf{PFS, Physics} \footnote{All initials are pseudonyms.} 
\end{displayquote} 
More so than content, the physics professors we spoke to emphasized the idea of the ``physics approach'' to science.
They did bring up content as an important component, but this approach seemed to be the underlying, foundational element in physics' relationship to other disciplines. 

Non-physics professors focused more on how their disciplines \textit{use} physics. 
In doing so, they articulated how physics functions as an important component of their discipline.
\begin{displayquote}
I think probably as biologists we would say: there is so much physics as an underpinning to the chemistry as an underpinning to the biology, that I kind of can't imagine a student understanding biology in the way I would like them to if they had had no physics.	\textbf{PAD, Biology}
\end{displayquote}
Elsewhere, non-physics professors provided numerous, specific examples of physics content that supports their disciplines. 
Furthermore, non-physics professors often mentioned scientific skills they viewed as central to the study of physics that they valued. 
The quote above summarizes and represents the views of non-physics professors; through a variety of ways, physics plays a key role in the complete science education of students. 

As opposed to the concrete and detailed ideas of the professors, students were more uncertain about the relationship between the physics and other disciplines. 
\begin{displayquote}
It can't all just be prep for Physical Chemistry ... I guess maybe like the problem solving	that's in physics, that's all I can really think of. That's all I think I will take away from this 	course, because I don't think I'll be using any kinematic equations ever again in my life.  \textbf{LY}
\end{displayquote}
This quote demonstrates the uncertainty of students when it comes to physics' interdisciplinary relationship. 
Still, this student is somewhat aligned with non-physics professors, especially molecular biology professors, in emphasizing the relevance of physics to physical chemistry. 
Problem-solving, too, emerged as a key scientific skill, mentioned and expanded upon by physics and non-physics professors alike. 

Nonetheless, students did express notions of interdisciplinary possibilities that were at odds with those expressed by professors:
\begin{displayquote}
I think [molecular biology] requires physics just cause [molecular biology] is already like between chemistry and biology. So it's just like looking at molecules. [The professor] did have some examples of ``this molecule is moving like this,'' but it's very specific examples. So it's not so much of physics can apply to so much of [molecular biology]. \textbf{VQ}
\end{displayquote}
Here, VQ's concept of physics' interdisciplinary role centers around motion, and so they restricted physics and molecular biology connections to ``very specific examples.''
Non-physics professors, as exemplified in PAD's quote, held an expansive view of physics' interdisciplinary relevance.
Importantly, this student was not taking a wild guess, but based their conjecture on their experiences in the course. 
The student's phrasing at the beginning of this quote, specifically their use of ``requires,'' reveals and represents how many students constructed their ideas of physics' interdisciplinary role: because medical schools and some life-science departments require physics, that must mean physics has an interdisciplinary role.

\mysubsection{Goals}
When asked what success would mean for them in this course, most students pointed towards interdisciplinary applications of their physics knowledge. 
For example, one student replied,
\begin{displayquote}
I would say being able to actually make the connection from all the other sciences that I'm taking to physics, because physics always seemed so non-life-sciences. ...So if I do see myself making more of those connections, if I'm like, ``this is how the physics of this concept will work.'' If it's a chemistry or biochem concept or even a neuro concept. I think being able to just make that connection automatically, I would say that I would have succeeded in physics then. \textbf{LM}
\end{displayquote}
The quote above speaks to a potential mismatch within student perceptions of this course:  students often articulated success in terms of interdisciplinary relevance, but also perceived physics as disconnected from their own discipline. 
It is also worth noting that LM is using the phrase ``life-sciences'' inclusively here, to include chemistry along with biochemistry and neuroscience. 
Here, we see further evidence of students' uncertainty about the interdisciplinary role of physics; this quote provides another example of what students think physics' interdisciplinary role might be.

The perception of a disconnection between physics and other disciplines existed among nearly all student interviewees, including those students who articulated goals related to the interdisciplinary role of physics. 
All the students who did not describe success in terms of interdisciplinary applications explained that they either did not believe or did not know if physics would be useful for them going forward. 

Students who did articulate goals related to the interdisciplinary role of physics were still uncertain about those goals. For one student, an interest in structural biology led their professor to recommend taking physics. Still, they said, 
\begin{displayquote} I feel like I still don't have the understanding of what I need exactly from physics. I guess in my head physics is always just like the physical representation of things and that's something that would obviously tie in with the structure of something.	\textbf{JE} 
\end{displayquote}
While students were consistently vague about how or where such applications would take place, non-physics professors provided some specifics. 
When asked how they would describe success in a physics course, one non-physics professor replied, 
\begin{displayquote}
It's really helpful if students do have a really intuitive understanding of electrical properties [...] So we start with a model circuit, with a real resistor and a capacitor, and then talk about the cellular analogs -- ion channels closing serve as resistors, and the thin cell membranes are good capacitors. And so, being able to take that basic information and then apply it to a cell, that's key. \textbf{PNK, Neuroscience}
\end{displayquote}
Represented in the quote above, non-physics professors see relevance and value in physics content knowledge as well as in non-content skills -- intuition, model-building, quantitative literacy and reasoning, qualitative or logical reasoning skills, and problem solving skills -- that are perceived as central to physics. 
Biology professors, in particular, noted that their introductory series does not focus much on quantitative skills and that they view physics as an ideal setting for their students to develop such skills. 

The physics professors we spoke to expressed a similar sentiment, but with a stronger emphasis on the non-content skills -- once again invoking the idea of a ``physics approach'' to science.
\begin{displayquote}
Really its role is to introduce to a population that will probably never take another physics course, how physicists go about approaching difficult problems, because that is the meta-skill that we're always trying to teach. You can call it the scientific method. It's probably a little bit finer than that because I think there are many types of scientists and I think physicists, well at least we like to think we're special in our approach. [...] I don't think they need to come away with the technicalities of a lot of physics, it's the approach that is more important to me. \textbf{PCS, Physics}
\end{displayquote}
Although this quote references the role of General Physics, which is discussed in the next section, it is relevant for this section as well because it outlines a key learning goal for the physics professors. 
Their descriptions of this ``physics approach'' overlap significantly with the non-content skills referenced above. 
For these physics professors, the ``physics approach'' they want students to practice includes breaking down an observed phenomenon into its fundamental components in order to generate a simple model. 
This process involves quantitative skills and reasoning as well as the intuition and confidence to make assumptions and test their models.

It is noteworthy that the non-physics professors brought up specific physics content more than the physics professors. 
The physics professors spoke about content in terms of exposure, appreciation, and the skills students learn through understanding concepts (tying it all back to the ``physics approach''). 
Non-physics professors, on the other hand, detailed how their courses and disciplines build on and use physics concepts. 
With regards to General Physics, they focused the conversation around preparation: 
\begin{displayquote}
I mean coming in we can't spend two weeks talking about waves in [Physical Chemistry], so having that background is really good. \textbf{PKI, Chemistry}
\end{displayquote}
This quote exemplifies a general trend of non-physics professors framing learning content knowledge as a valuable outcome of General Physics.

\mysubsection{Course Role/Function}
Every student interviewee referenced requirements of some kind as a factor in their enrollment in General Physics. 
MCAT preparation and medical school requirements influenced most students, and other examples included satisfying the physics pre-requisite for the Physical Chemistry course, and requirements for other graduate programs (e.g. architecture). 
For many students, requirements were the primary reason for their enrollment.
\begin{displayquote}
Well it's required for pre-health. I mean that was the main reason. I do like physics but I don't know if I would have taken it if it wasn't required. \textbf{VH}
\end{displayquote}
Non-physics professors, too, emphasized the relevance of such requirements and connected them to the interdisciplinary role of physics and learning outcomes in the course:
\begin{displayquote}
Because the thing that people are looking for in graduate students in neuroscience are strong quantitative skills. \textbf{PEK, Neuroscience}
\end{displayquote}
Non-physics professors clarified that some graduate programs may not explicitly require physics, but the importance of the aforementioned non-content skills leads professors to strongly recommend that students take physics. 
In effect, there exist informal requirements that lead students to take General Physics.

For both the graduate school requirements and the department requirements, non-physics professors framed the role of General Physics as preparation:
\begin{displayquote}
The reason for the requirement for [General Physics] is that it provides the foundation for Physical Chemistry. \textbf{PKI, Chemistry}
\end{displayquote}
Even though this quote is specific to the Physical Chemistry course, it does capture a general theme among non-physics professors: physics provides a foundation of knowledge that is relied and built upon in advanced non-physics courses and research settings.

As mentioned previously, the requirements lead students to expect interdisciplinary relevance in the course. 
Non-physics professors substantiate this reasoning, and explain that the requirements come from the important interdisciplinary role that physics occupies. 
Given that role, non-physics professors expressed a desire for the course to emphasize interdisciplinary connections, applications, and overall relevance. 

Here, the physics professors we spoke to bring up concerns. 
First of all, addressing the scope of interdisciplinarity in the course, the breadth of non-physics investment is non-trivial.
\begin{displayquote}
I don't think it makes sense to speak of a role rather than roles, because there's so many. There's so many different interests from different directions in the class. \textbf{PFS, Physics}
\end{displayquote}
There exists significant variation among the role of this course based on the discipline of the students who take it, especially given that the course aims to serve not just life-science disciplines but also other fields, e.g. chemistry, geology, and environmental analysis.
Not only the relevant content, but also the interdisciplinary applications, vary across all these different disciplines. 

Given both this breadth of interdisciplinary investment in the course, as well as concerns about inauthentic interdisciplinary teaching, the other physics professor came to a different conclusion than the non-physics professors about the role of General Physics.
\begin{displayquote}
I did feel like at the end of the day it was about introducing physics and physics for its own self and not trying to justify physics. I think sometimes that actually is a negative too, it comes across as defensive, like ``oh physics is important, see?'' I think if you teach it right, it becomes kind of obvious in and of itself. You're like, ``oh, this is interesting.'' That was what I wanted, I just wanted them to be interested and to be engaged.  \textbf{PCS, Physics}
\end{displayquote}
This physics professor believes in the unique value of a physics education -- as discussed elsewhere in their interview, ``physics for its own self'' refers to many of the same skills non-physics professors mentioned (i.e. the ``physics approach'' as described in the quote from PCS in section IV.B). 
Again, the physics professors we spoke to shared the view of their non-physics colleagues, framing physics as a unique and important component of a student's science education. 
Furthermore, in PCS's experience, justifying physics refers to inauthentic attempts at interdisciplinary pedagogy that distracts from the goals of the course. 
\begin{displayquote}
There was definitely a sense that the biology that was included in it previously often looks to be pandering and wasn't genuine. And they saw through that. And when students feel that they're being pandered to, I think you lose a lot of the credibility. I think just the genuineness of the class goes down, and it suffers. \textbf{PCS, Physics}
\end{displayquote}
For the physics professors we spoke to, an emphasis on interdisciplinary science can obscure the parts of physics they feel to be most valuable. 
So, their hesitance to focus on interdisciplinary connections and applications comes from a recognition of the distinct interdisciplinary role that physics, and this course, plays. 

\mysection{Discussion}\label{discussionSection}
The quotes above were specifically chosen because they represent and effectively summarize the views of the different stakeholder populations.
We see general agreement about the interdisciplinary role of the General Physics course and with regards to some specific aspects of why physics is important. 
General Physics was described as a distinct and valuable component of a student's science education, and there is significant alignment between non-physics and physics professors on the subject of physics' interdisciplinary value. 

Non-physics and physics professors provide generally aligned, though also distinct, perspectives on the interdisciplinary role of physics. 
Non-physics faculty are especially well-suited to identify relevant topics and concepts from physics, and in their interviews physics professors acknowledged the value of such specific input. 
Physics professors' views on curriculum content primarily reflects their prioritization of non-content skills:  they want students to develop a physical intuition and recognize how physicists quantify and model physical phenomena. 
Overall, the non-content skills mentioned by physics and non-physics faculty include physical intuition or ``common sense,'' comfort with model-building, quantitative literacy and reasoning, qualitative or logical reasoning skills, problem solving skills, and confidence to use their knowledge and skills.  
Physics professors described these skills as part of a ``physics approach'' to science. Non-physics faculty viewed these skills as unique and important learning outcomes of a physics course just as much as physics content.

Despite the alignment with regards to the interdisciplinary importance of physics, professors presented different implications for the teaching of General Physics. 
Non-physics professors connected their emphasis on specific content to a more interdisciplinary approach to teaching the course -- both teaching relevant physics concepts and highlighting interdisciplinary connections and applications. 
Students expressed a similar sentiment: they expect and are eager to see the relevance of what they are learning.
Physics professors expressed concerns about inauthentic attempts to incorporate non-physics material and ``justify physics.'' 
These concerns are encapsulated in the following quote from physics professor PCS:  ``I almost felt like trying to modify the content based on that audience could go wrong more than it could go right.''  

We can understand this divide as it relates to the professors' differing descriptions of physics' interdisciplinary value. 
For the physics professors we spoke to, the value centers around the ``physics approach.'' 
Accordingly, they expressed concerns about how inauthentic attempts at interdisciplinary pedagogy might prevent students from developing and practicing such an approach. 
With the breadth of non-physics investment, too, a key commonality among the different disciplines was the value placed on this ``physics approach.''
Non-physics professors primarily framed the interdisciplinary value of physics through how their disciplines and how their students might \textit{use} physics knowledge and associated skills.
Accordingly, non-physics professors emphasized the importance of various scientific skills associated with physics, and, more so than the physics professors, they highlighted numerous concepts and topics in physics that are key material for their discipline. 
Furthermore, they expressed that explicitly highlighting interdisciplinary connections and applications would best prepare students to build on their physics content knowledge in non-physics settings.

Despite this distinction in the perspectives of the stakeholders, each of them aligns with prior research involving IPLS courses.
Our students' emphasis on interdisciplinary relevance aligns with findings about the role of authentic interdisciplinary connections as sources of interest and engagement \cite{geller_sources_2018}. 
The non-content skills mentioned above bear a striking resemblance to the new MCAT competencies \cite{hilborn_physics_2014}, and the epistemological goals of the University of Maryland, University of Minnesota, and Swarthmore courses \cite{redish_reinventing_2009, redish_nexus/physics:_2014, crouch_introductory_2014}. 
Several relevant physical topics mentioned by non-physics professors, such as electricity, fluids, radiation, (geometric) optics, also appear in the literature \cite{hilborn_physics_2014, redish_nexus/physics:_2014, crouch_introductory_2014}.
At the same time, the concerns of physics professors documented here also align with prior IPLS research:  inauthentic interdisciplinary connections are not sources of interest and engagement \cite{geller_sources_2018}.

\section{Conclusions and Future Work}\label{conclusionSection}
The findings presented here answer our original research question: How are the goals of the different stakeholders (students, physics faculty, non-physics faculty/departments) aligned? How are they distinct? 
Stakeholder perspectives generally align with regards to the interdisciplinary role of physics.
Professors, in particular, had nearly exact alignment with the scientific skills they described as essential to physics.
Despite some overall uncertainty amongst students about how physics might be relevant, they too focused on interdisciplinary relevance as a goal for this course.
Across most of our interviewees, their notion of the current role of General Physics represents such overall alignment towards a single goal: to communicate the unique value of physics to students so they can build off of their knowledge and apply their skills in future courses, research, and/or careers.

Furthermore, the views of our stakeholders align with prior IPLS research. 
It is notable that these themes continue to be salient among a broader range of non-physics professors.
That said, it is also important to note that this project did not seek to produce a comprehensive list of relevant topics, nor did this project solicit feedback from geology or environmental analysis professors.
Given the unique context of General Physics, it is imperative that future research identify relevant topics for the course rather than rely solely on prior IPLS research. 

The primary distinction between stakeholder perspectives comes from their views on the implications for the course.
Where students and non-physics professors expressed desires for the active presence of interdisciplinary applications and connections in General Physics, physics professors identified potential risks implementing such interdisciplinarity. 
Through reconciling these concerns with the perspectives of non-physics professors and students, 
we can address these distinctions and create a course that better addresses the needs of all stakeholders.

This project provides some insight into which topics and concepts are relevant for students in General Physics.
It is in line with the role of General Physics as we have described it here to alter some of the content in the current curriculum, but our data does not have a definitive implication for the form of such changes.
The general alignment of our results with IPLS research in spite of the broader interdisciplinary population of this course suggests that future research concerning this course should, like this project, adapt the processes of IPLS research to the Pomona context.
For example, our data strongly suggest that authentic interdisciplinary gestures (examples, applications, and connections) serve as sources of engagement and interest, but as Geller et al. \cite{geller_sources_2018} and Nair and Sawtelle \cite{nair_operationalizing_2019} remind us, engagement, interest, and relevance are complex constructs with varied meanings. 
Not only would adapting such IPLS research to the General Physics context provide useful information to support research-based reformation of this course, but also, given the centrality of physics-biology as their interdisciplinary context, it is worth investigating how the results of  \cite{geller_sources_2018, nair_operationalizing_2019} as well as \cite{ geller_bridging_2019,  sawtelle_leveraging_2016, gouvea_framework_2013} transfer and compare to different interdisciplinary contexts.

In summary, our results fit into the paradigm of interdisciplinary education established by prior IPLS research, despite the ways in which the course context of General Physics departs from traditional IPLS settings.
Our research outlined the interdisciplinary value of physics and presented perspectives on the implications of this value for the teaching of General Physics.
Moreover, given the wide variety of contexts in which physics for non-physics courses are taught, the process for obtaining and analyzing our data -- our approach for collaborating with non-physics stakeholders -- is of equal importance. 
By seeking perspectives from a wide range of stakeholders, both students and faculty, across a variety of disciplines, we can discover how to improve physics for non-physics-major courses, and identify what makes them unique settings for physics learning.

\mysectionNoNumber{Acknowledgements}
We thank the faculty and staff of the Pomona Physics Department for their consistent and enthusiastic support of this project, especially Elijah Quetin. 
We also thank Makaela Stephens for her advice and insight regarding qualitative research, as well as Jessica Hoehn, Joel Corbo, and Gina Quan for their thoughts and pointers around qualitative analysis methodologies.
We thank Benjamin Geller for feedback on a version of this manuscript.
We also gratefully acknowledge the Pomona College Physics Department for providing funding to support this research project.
Finally, we thank the interview participants for their willingness to share their experiences, perspectives, and insights; without them, this research would not be possible.

\pagebreak

\bibliographystyle{apsrev} 
\bibliography{references.bib} 

\end{document}